\documentclass[conference]{IEEEtran}
\IEEEoverridecommandlockouts
\usepackage{cite}
\usepackage{amsmath,amssymb,amsfonts}
\usepackage{algorithmic}
\usepackage{graphicx}
\usepackage{textcomp}
\usepackage{xcolor}
\usepackage{float}
\usepackage{tabularx}
\usepackage{booktabs}
\def\BibTeX{{\rm B\kern-.05em{\sc i\kern-.025em b}\kern-.08em
    T\kern-.1667em\lower.7ex\hbox{E}\kern-.125emX}}

\makeatletter
\newcommand{\linebreakand}{%
    \end{@IEEEauthorhalign}
    \hfill\mbox{}\par
    \mbox{}\hfill\begin{@IEEEauthorhalign}
}

\begin{document}
\title{A Comparative Analysis of Instagram and TikTok as Islamic da'wahh Media in the Digital Era\\}

\author{\IEEEauthorblockN{1\textsuperscript{st} Wisnu Uriawan}
\IEEEauthorblockA{\textit{Informatics Department}\\
\textit{UIN Sunan Gunung Djati Bandung}\\
Jawa Barat, Indonesia\\
wisnu\_u@uinsgd.ac.id}
\and
\IEEEauthorblockN{2\textsuperscript{nd} Muhammad Saifurridwani 'Ijazi}
\IEEEauthorblockA{\textit{Informatics Department}\\
\textit{UIN Sunan Gunung Djati Bandung}\\
Jawa Barat, Indonesia\\
m.saifurridwani@gmail.com}
\and
\IEEEauthorblockN{3\textsuperscript{rd} Nizzami Ramdhan Arraudy}
\IEEEauthorblockA{\textit{Informatics Department}\\
\textit{UIN Sunan Gunung Djati Bandung}\\
Jawa Barat, Indonesia\\
nizzamiramdhan@gmail.com}
\linebreakand
\IEEEauthorblockN{4\textsuperscript{th} Onixa Shafa Putri Wibowo}
\IEEEauthorblockA{\textit{Informatics Department}\\
\textit{UIN Sunan Gunung Djati Bandung}\\
Jawa Barat, Indonesia\\
onixashafa@gmail.com}
\and
\IEEEauthorblockN{5\textsuperscript{th} Radithya Dwi Santoso}
\IEEEauthorblockA{\textit{Informatics Department}\\
\textit{UIN Sunan Gunung Djati Bandung}\\
Jawa Barat, Indonesia\\
radithyadwisantoso78@gmail.com}
}

\maketitle

\begin{abstract}
This research aims to analyze and compare the effectiveness of Islamic da'wahh on two popular social media platforms, namely Instagram and TikTok. The analysis is conducted based on four main aspects: media characteristics, da'wahh communication strategies, audience engagement effectiveness, and user behavioral responses. The research employs a a mixed-methods approach integrating qualitative content analysis and descriptive quantitative metrics through observations of the popular da'wahh account @hananattakistory during the period of October to November 2025. The findings indicate that TikTok excels in the effectiveness of disseminating da'wahh messages through high interaction rates, achieving an engagement rate of 1.42\%. In contrast, Instagram records a higher total interaction with an engagement rate of 5.47\%, reflecting deeper and more reflective audience involvement. Qualitative analysis shows that TikTok is more effective in the initial stage of capturing audience attention (awareness), while Instagram is stronger in building loyalty and fostering a digital da'wahh community. Thus, combining the use of both platforms can serve as a complementary digital da'wahh strategy, expanding reach while deepening the understanding of Islamic values in the era of social media. The main contribution of this study lies in its comparative, empirically grounded approach that integrates both qualitative and quantitative analyses to map the effectiveness of da'wahh across platforms, providing a strategic foundation for developing adaptive and contextual digital da'wahh for young Muslim audiences.
\end{abstract}


\section{Introduction} \label{sec:introduction}

The development of information and communication technology in the digital era has brought significant changes to patterns of social interaction and information dissemination worldwide. One of the most prominent impacts is the transformation of how society communicates through social media, which has now become the most dynamic virtual public space. Platforms such as Instagram and TikTok have become an integral part of the lives of millennials and Gen Z, enabling rapid, visual, and interactive information exchange \cite{7}.

TikTok is a short-form video platform that highlights visual creativity and personalized content algorithms, while Instagram focuses on visual aesthetics through photos and videos with interactive features such as Reels, Stories, and IGTV. Both have become the primary media for young users to interact, express themselves, and consume various forms of information. This phenomenon aligns with data showing that Generation Z and millennials allocate most of their time to interacting on social media, making it the main space for information and educational consumption \cite{10}. In Indonesia, about 86.6\% of active internet users engage with social media, with TikTok usage among Gen Z reaching 76\%, surpassing YouTube at only 71\% \cite{10}. This data illustrates the dominant role of TikTok among young generations, making it a potential medium for spreading religious values and messages.

\begin{figure}[H]
\centering
\includegraphics[width=0.45\textwidth]{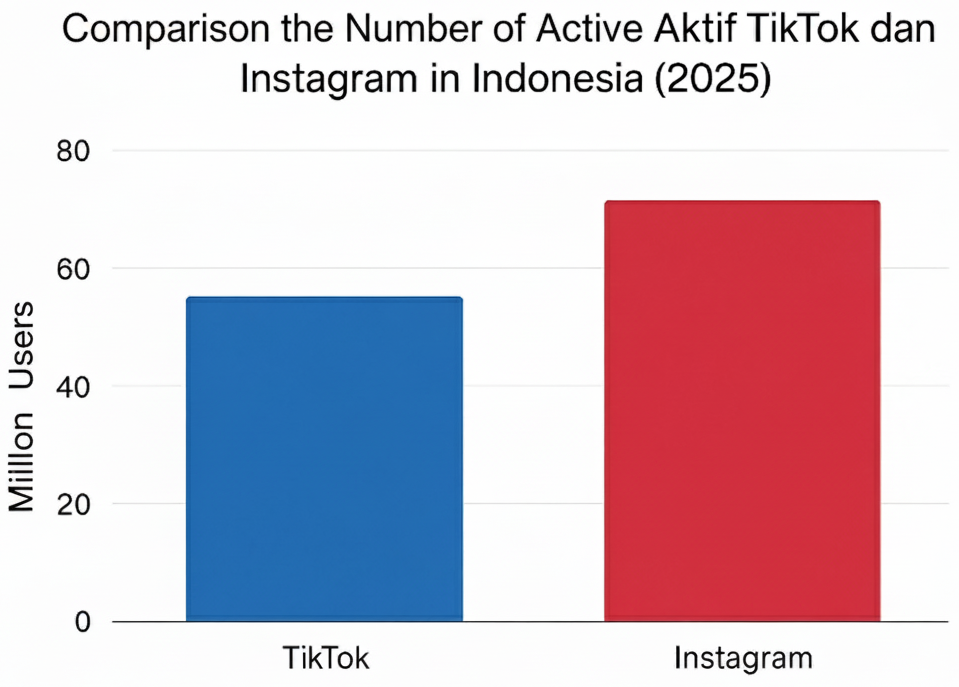}
\caption{Comparison of Active TikTok and Instagram Users in Indonesia (2025)}
\label{fig:user_comparison}
\end{figure}

The phenomenon of high TikTok and Instagram penetration, as shown in Figure \ref{fig:user_comparison}, strengthens the position of both platforms as dominant media in the interaction and content consumption patterns of young generations. This condition presents strategic opportunities for the more effective and massive dissemination of Islamic da'wahh in digital spaces.

These changes also affect various aspects of life, including Islamic da'wahh, where methods of conveying religious messages are no longer limited to face-to-face sermons but have expanded to creative and easily accessible digital content \cite{6}. Instagram and TikTok have transformed into dominant da'wahh arenas, especially among millennials and Gen Z, who are the primary users of these platforms \cite{1,2}. da'wahh has shifted from conventional pulpits to concise, visually appealing audio-visual formats that rely on social media algorithms \cite{3}. This phenomenon represents a new form of religious communication that is more adaptive to technological developments. Islamic messages can now be packaged in short videos, infographics, or visual posts on platforms such as Instagram and TikTok \cite{7}, making them relevant and efficient da'wahh tools in the digital era because they can transcend spatial and temporal boundaries at relatively low cost \cite{8}. 

As two platforms with distinct characteristics and algorithms, TikTok and Instagram offer unique approaches to disseminating da'wahh messages. TikTok offers short video formats with interaction-based algorithms that allow da'wahh content to spread widely even without direct user searches \cite{9}. A distinctive feature that differentiates TikTok from other social media platforms is the For You Page (FYP) mechanism, an algorithm-driven system that distributes content based on user interactions rather than follower relationships. This mechanism enables da'wahh content to reach broader audiences organically, making TikTok highly effective in increasing initial exposure and awareness of Islamic messages. Meanwhile, Instagram provides flexibility through features such as Reels, Stories, and IGTV, allowing preachers to present messages in visually appealing ways \cite{9,10}. These platforms have different characteristics but are equally influential in shaping information consumption behavior and religious expression among young generations. This opens vast opportunities for da'i and Islamic content creators to reach wider audiences, especially young users who are more responsive to visual communication styles. However, this dynamic also raises questions about the effectiveness and quality of digital da'wahh messages, as well as how essential Islamic values can be preserved amid fast-paced content flows \cite{5}.

Beyond algorithms and content formats, interactivity becomes another advantage of both platforms. TikTok and Instagram enable direct communication between preachers and audiences through comment sections, direct messages, and live streaming sessions \cite{9}. This dialogic communication creates more personal connections, where audiences may ask questions, engage in discussions, or even receive direct guidance. Interestingly, many young users engaged in digital da'wahh on TikTok hold strong religious motivations. They are driven by the desire to gain reward and share religious knowledge. Some even believe they can earn ongoing reward (\textit{jariyah}) simply by reposting da'wahh content from other creators. This shows that social media functions not only as entertainment but also as a spiritual and educational medium for young Muslim audiences.

After observing the dynamics of digital da'wahh at the user and creator levels, it is also important to understand how this phenomenon evolves within the global context. da'wahh through social media has proven to be an essential instrument for strengthening religious literacy and countering the spread of extremism or Islamophobia. Platforms such as TikTok and Instagram provide open dialogic spaces that allow real-time interaction between preachers and audiences. According to Taufikurrahman and Setyowati \cite{7}, these digital platforms enhance the effectiveness and efficiency of da'wahh, but at the same time require ethical responsibility and digital competence from preachers. In Indonesia, the phenomenon of digital da'wahh on TikTok and Instagram has been growing rapidly. Abdulhalim et al. \cite{4} found that da'wahh content on TikTok successfully attracts positive responses from young audiences because it is packaged in visually appealing styles while still adhering to Islamic principles. Meanwhile, Parhan et al. \cite{5} emphasize that TikTok has significant potential as a da'wahh medium among Muslim university students, although there remain risks of content misuse and challenges in maintaining user religiosity.

Considering these phenomena, it becomes essential to analyze in depth how Instagram and TikTok are used as Islamic da'wahh tools in the digital era, especially in the context of Indonesian youth who are active users of both platforms. Although digital da'wahh on Instagram and TikTok offers great potential for reaching young audiences, research gaps remain that require further investigation. This study focuses on comparatively analyzing the effectiveness, communication strategies, and interactivity levels of Islamic da'wahh on Instagram and TikTok. This analysis is expected to provide a comprehensive understanding of digital da'wahh communication patterns, the effectiveness of religious message delivery, and its implications for strengthening Islamic values in modern society. Therefore, this research aims to explore strategies, challenges, and the impacts of using social media as a relevant, moderate, and inclusive da'wahh instrument in the era of digital transformation.

Based on the above discussion, this study offers novelty in both approach and research focus. Unlike previous studies that generally examined a single digital da'wahh platform separately or merely reviewed social media effectiveness in general, this study presents a comparative, empirically grounded analysis of two major platforms: TikTok and Instagram. By integrating qualitative and quantitative approaches, this research seeks to identify differences in communication strategies, audience interaction patterns, and the effectiveness of delivering Islamic da'wahh messages across both platforms. Thus, this study provides original contributions to the development of digital da'wahh studies, particularly in the context of utilizing social media algorithms and the characteristics of young Muslim generations in Indonesia. The results are expected to serve as a foundation for formulating more effective, moderate, and adaptive digital da'wahh strategies in response to contemporary social media dynamics.

\section{Related Work} \label{sec:related-work}
The use of social media as a medium for Islamic da’wah has been extensively studied. Research indicates that the development of digital technology has brought significant changes to the patterns of Islamic da’wah. Through a Systematic Literature Review (SLR) of 3,480 publications spanning 2013–2023, it was found that social media platforms such as YouTube, TikTok, Instagram, and Facebook have become effective channels for disseminating religious messages, particularly to millennials and Generation Z. These studies confirm that the effectiveness of da’wah on social media is heavily influenced by the quality of the da’i (preacher), the packaging of the message, as well as the frequency and intensity of content delivery. Moreover, features such as video uploads, live streaming, and podcasts enable da’i to interact directly with their audience, thereby making da’wah more interactive and accessible \cite{16}.
However, before delving into a detailed discussion of the specific platforms Instagram and TikTok, it is essential to first map the broader paradigm shift in da’wah itself resulting from the digital revolution. Table \ref{tab:perbedaan_dakwah} below summarizes the fundamental differences between the traditional da’wah approach (pre-2010 era) and the contemporary digital da’i approach (2020–present).

\begin{table}[htbp]
\centering
\caption{Differences Between Traditional Preaching Approaches and Contemporary Digital Preaching}
\label{tab:perbedaan_dakwah}
\begin{tabularx}{\linewidth}{c X X}
\toprule
\textbf{No} & \textbf{Traditional Preachers (Old)} & \textbf{Digital Preachers (New)} \\
\midrule
1 & Preaching conducted in mosques or religious study assemblies & Preaching conducted anytime and anywhere (home, street, mini studio) \\
2 & The number of preachers was limited and highly selective & The number of preachers is very large and diverse \\
3 & Content focused on general themes such as enjoining good and forbidding wrong & Contextual content aligned with youth issues and contemporary daily realities \\
4 & Limited space for conveying opinions & Freedom to deliver messages through various social media platforms \\
5 & Preaching reach limited to oral or written communication & Extremely wide reach enabled by platform technologies and algorithms \\
6 & Classical sermons as the primary method & Varied methods: short videos, memes, storytelling, challenges, live sessions, duets, podcasts, etc. \\
\bottomrule
\end{tabularx}
\end{table}

The study entitled “The Digital Age of Religious Communication” explains that social media has fundamentally transformed the patterns of religious communication. Platforms such as Instagram and TikTok have enabled the democratization of religious message dissemination, allowing individuals to act as agents of da’wah without institutional constraints. However, it also highlights challenges such as message fragmentation and the phenomenon of information overload, which can obscure the deeper meaning of religious teachings. These findings are highly relevant to the context of the present study, which comparatively analyzes the effectiveness of Islamic da’wah on these two popular platforms \cite{20}.

Several follow-up studies have also highlighted the specific roles of each platform. Instagram is considered effective as a medium for da’wah due to its ease of use and strong visual appeal in conveying Islamic messages \cite{17}. Meanwhile, TikTok has proven capable of enhancing religious values through creative and communicative short-form videos \cite{18}.
Building on these findings, the present study seeks to extend previous research by focusing on a comparative analysis of the effectiveness of Instagram and TikTok as platforms for Islamic da’wah in the digital era. The aim is to understand the extent to which each platform is able to foster audience engagement and acceptance of da’wah messages among users.

The use of social media as a medium for Islamic da’wah has been widely studied in recent years. Several studies show that digital platforms such as YouTube and Instagram offer significant opportunities for preachers (da’i) to disseminate Islamic teachings in creative ways. However, challenges remain, particularly related to low technological literacy and the potential spread of inaccurate information \cite{11}.
Other studies confirm that since 2019, TikTok, Instagram, YouTube, and Facebook have become increasingly effective as da’wah media, with their success largely depending on the quality of the da’i, the packaging of the message, and the suitability of the delivery method to the characteristics of millennials and Generation Z \cite{12}.

Furthermore, several studies have highlighted the role of digital da’wah on TikTok and Instagram in shaping the political preferences of young Muslim generations. Drawing on Gramsci’s theory of hegemony, one such study underscores the importance of digital literacy and a dialogic da’wah approach to prevent religious content from triggering polarization \cite{13}. On the other hand, a semiotic analysis of da’wah messages in Husain Basyaiban’s TikTok content, employing Roland Barthes’ framework, reveals that visually and communicatively packaged da’wah messages successfully instill values of tolerance, morality, and social responsibility among the younger generation \cite{14}.

Additionally, previous research focused specifically on Instagram has identified a significant correlation between video views and the number of likes on Islamic da’wah content. By utilizing web-scraping technology and Pearson correlation analysis, the study evaluated user engagement levels and concluded that such analytics can assist preachers in better understanding audience preferences \cite{15}. In contrast to these works, the present study seeks to compare two highly popular platforms—Instagram and TikTok—to examine differences in their effectiveness as contemporary da’wah media.

Overall, prior studies consistently demonstrate that social media, particularly TikTok and Instagram, possess enormous potential as creative, interactive, and highly relevant da’wah instruments for the digital-native generation. Nevertheless, their success remains heavily dependent on digital literacy, ethical communication practices, and the da’i’s ability to package messages in a moderate and contextually appropriate manner.

Although numerous previous studies have explored the effectiveness of social media as a medium for Islamic da’wah, most have either focused on a single platform in isolation or examined social media in general without conducting a detailed comparative analysis of the distinct characteristics and da’wah strategies across different platforms. Therefore, this study aims to fill that research gap by conducting a comparative analysis of the effectiveness, communication strategies, and interactivity levels of Islamic da’wah on Instagram and TikTok. It is expected that the findings will provide a more comprehensive understanding of digital da’wah communication patterns, the effectiveness of religious message delivery, and their broader implications for reinforcing Islamic values in modern society. Ultimately, this research seeks to explore the strategies, challenges, and impacts of using social media as a relevant, moderate, and inclusive instrument of da’wah in the era of digital transformation.

\section{Methodology} \label{sec:methodology}

\subsection{Research Type and Approach}

This study employs a qualitative-comparative approach supported by descriptive quantitative data within a mixed-methods design framework. This strategy was selected because the research not only examines the meaning and delivery strategies of da’wah but also objectively and measurably compares the effectiveness of two digital platforms—Instagram and TikTok.

The qualitative approach is used to explore the meanings, messages, discursive forms, and communication patterns employed by da’wah content creators on each platform. Meanwhile, quantitative data were directly obtained through digital observation of content interaction metrics (views, likes, comments, and shares) extracted from the public statistics features of selected accounts on TikTok and Instagram. These data were manually recorded into a digital observation spreadsheet containing variables such as posting date, content type, follower count, and cumulative total interactions.

Quantitative data collection was conducted periodically throughout the research period (October–November 2025), with recordings taken every two days to ensure accuracy given the highly fluctuating nature of engagement metrics. The resulting numerical values were subsequently processed using the Engagement Rate formula (see Equation \ref{eq1}) and presented in comparative tables and graphs across platforms. These quantitative data serve to support qualitative findings with measurable empirical evidence, thereby enabling a more valid and transparent comparative analysis of the two platforms.

This research model aligns with contemporary digital media studies that advocate integration between interpretative and data-driven approaches \cite{19}. Consequently, the analysis yields not only interpretive descriptions but also verifiable results based on digital performance indicators.

The primary strength of the mixed-methods sequential explanatory design lies in its capacity for complementary data triangulation: qualitative findings concerning meaning, linguistic style, and rhetorical strategies of da’wah are validated and enriched by quantitative evidence of real-world audience reception. The selection of the October–November 2025 period was deliberate; these months coincide with the commemoration of Mawlid al-Nabī (the Prophet Muhammad’s birthday, peace be upon him) and the beginning of the Islamic new year, periods that consistently trigger surges in da’wah content on both platforms. This timing thus allows the researcher to capture interaction patterns during a representative “high-engagement season.”

The decision to record data every two days was informed by preliminary trials showing that daily metrics on TikTok can fluctuate by 15–30\% due to the For You Page algorithm, whereas Instagram exhibits greater stability yet still requires intensive monitoring to capture the snowball effects of Reels and Stories features. Accordingly, this approach not only enhances the reliability of quantitative data but also enables the identification of temporal correlations between shifts in creators’ strategies (e.g., switching from short videos to live sessions) and corresponding spikes or declines in engagement rates. Ultimately, these insights strengthen the comparative argument regarding which platform demonstrates superiority in delivering da’wah messages to Generation Z and late millennials.

\begin{figure}[H]
\centering
\includegraphics[width=0.2\textwidth]{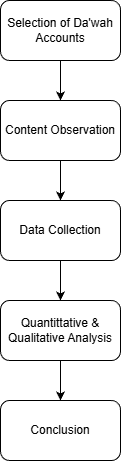}
\caption{The Flow of Stages of Digital Preaching Research on Social Media}
\label{fig:research_flow}
\end{figure}

As illustrated in Figure \ref{fig:research_flow}, this study adopts a sequential explanatory design consisting of five main stages:
(1) selection and mapping of targeted da’wah accounts based on criteria such as follower count, posting consistency, and content relevance to contemporary da’wah themes;
(2) intensive content observation over a full two-month period;
(3) parallel data collection, comprising real-time extraction of quantitative metrics and simultaneous archiving of content for qualitative analysis;
(4) integrated analysis employing thematic content analysis (qualitative) and descriptive-comparative statistical analysis (quantitative); and
(5) synthesis of findings, drawing of comparative conclusions, and formulation of strategic recommendations.
This sequential approach allows quantitative data to serve as an initial framework for identifying high-performing content, which is subsequently subjected to in-depth qualitative analysis to uncover the underlying factors driving engagement levels. Accordingly, each stage reinforces the others and is rigorously controlled through periodic checking and re-checking procedures, thereby enhancing the study’s internal validity, external validity, and overall reliability.

\subsection{Research Focus and Units of Analysis}

The focus of this study is directed toward a comparative analysis of the effectiveness of Islamic da’wah on Instagram and TikTok. Effectiveness is evaluated across three primary dimensions:
\begin{enumerate}
    \item Media Characteristics
    
    This dimension encompasses content formats and presentation styles (short videos, carousels, Reels, Stories, live streaming, IGTV), duration, use of background music, text overlays, transitions, as well as the algorithmic features of each platform (Instagram’s Explore and Reels algorithms versus TikTok’s For You Page) that significantly influence the visibility and distribution of da’wah content.
    
    \item Da’wa Communication Strategies
    
    This includes an examination of message delivery approaches (sermons, advice, inspirational stories, religious humour, challenges, duets, or stitches), the frequency and consistency of Qur’anic verses and hadith citations, rhetorical styles (direct preaching, storytelling, dialogic, or entertainment-edutainment), tone of communication (authoritative, friendly, humorous, emotional), narrative structure, and techniques for building emotional and cognitive closeness with the audience (parasocial interaction).
    
    \item Media Effectiveness
    
    Effectiveness is measured quantitatively through Engagement Rate by Reach (ERR), Average Engagement Rate (AER), applause rate, and amplification rate, and qualitatively through the depth and quality of discursive interactions (e.g., fiqh-related questions in comments, testimonials of behavioural change, light theological debates, or simple emotional support). The comparison aims to determine which platform excels in generating reach, resonance, and reaction to da’wah messages.
\end{enumerate}

The units of analysis in this study are multi-level and comprise not only the da’wah accounts themselves but, more importantly, individual pieces of content posted on those accounts. Each post possesses unique technical, narrative, and performative characteristics that reflect the specific communication strategies employed on the respective platform.

A total of approximately 120 content units were selected for in-depth comparative analysis. The dataset consisted of around 30 Instagram posts and 30 TikTok posts from the account @hananattakistory, supplemented by additional randomly sampled content to ensure the inclusion of diverse themes, styles, and interaction patterns. This sampling approach provides a balanced representation of both high-engagement content and everyday da’wah posts, enabling a more accurate comparison of platform-specific characteristics.
The multi-layered observation—combining format, theme, delivery technique, religious references, and interaction metrics—supports a nuanced analysis of how Instagram and TikTok shape the creation and reception of contemporary digital da’wah messages.

\subsection{Research Population and Sample}

The research population comprises all active Islamic da’wah accounts on TikTok and Instagram in Indonesia. Since analyzing the entire population is not feasible, purposive sampling was employed, whereby subjects were selected based on specific criteria that align with the research objectives.

The sample selection criteria are as follows:
\begin{enumerate}
    \item Accounts that are active and regularly upload da’wah content;
    \item Accounts exhibiting significant interaction/engagement levels;
    \item Accounts whose primary audience consists of Generation Z and millennials;
    \item Accounts that consistently present da’wah-oriented content;
    \item Accounts employing creative delivery styles while adhering to da’wah ethics.
\end{enumerate}

From this population, one Instagram account and one TikTok account were selected as the primary objects of study. Approximately 30 da’wa posts were chosen from each account for in-depth analysis, resulting in a total of approximately 120 content units—an amount deemed sufficient to provide a comprehensive picture of trends and patterns in contemporary digital da’wa.

Based on the above criteria, the popular da’wah account @hananattakistory was selected as the main observation target on both Instagram and TikTok. This account was chosen due to its large follower base, high posting frequency, consistent and highly relevant da’wah content for young Indonesian Muslims, and its representative adaptive, communicative, and visually oriented digital da’wah style. These characteristics make it an ideal case for achieving the study’s objective of comparing the effectiveness of the two platforms in disseminating Islamic messages.

Furthermore, the research deliberately targets Generation Z and millennial audiences, as these cohorts constitute the largest and most active user groups of TikTok and Instagram in Indonesia and represent the primary demographic of contemporary digital da’wah. Their media consumption patterns are heavily shaped by short-form video formats, visual aesthetics, and recommendation algorithms that prioritise engagement-driven content. Selecting accounts whose audiences are dominated by these younger generations therefore ensures that the analysed data accurately reflect current dynamics of digital da’wah within Indonesia’s social media ecosystem.

The two-month observation period (October–November 2025) was intentionally chosen to yield stable and representative data while capturing natural variations in posting behaviour and audience interaction. This duration allows the researcher to document algorithmic fluctuations, evolving content trends, and organic weekly engagement patterns. A shorter period risks bias from temporary viral spikes or anomalous interactions, whereas an excessively long period would generate an unmanageable volume of data disproportionate to the research design. Thus, the two-month timeframe strikes an optimal balance between observational depth, data reliability, and practical time constraints.

To further enhance representativeness and mitigate potential bias arising from reliance on a single creator, future extensions of this study may incorporate additional high-performing accounts for cross-validation. Nevertheless, the focused selection of @hananattakistory—combined with rigorous multi-dimensional analysis—provides a robust foundation for generating valid comparative insights into platform-specific da’wah effectiveness.

\subsection{Research Period} 

Data collection was carried out over a two-month period (October to November 2025). This timeframe was deliberately chosen to capture organic content dynamics that naturally follow the posting cycles of creators and the behavioural patterns of their audiences. The relatively extended duration also allows for a more accurate assessment of interaction stability, message consistency, and long-term da’wah effectiveness, while remaining manageable within the scope of the study. Moreover, this period coincides with key Islamic commemorative events—particularly the Mawlid season and the approach of the new Islamic year (Muharram)—which typically trigger a significant surge in religious content across both platforms, thereby providing a rich, high-engagement context that is highly representative of peak da’wah activity in Indonesia.

\subsection{Data Collection Techniques}

Data were collected through three primary techniques: digital observation, coding instruments, and secondary documentation. The entire process was conducted simultaneously on Instagram and TikTok over the two-month research period, enabling the researcher to capture authentic, real-time da’wah activities in their natural digital environment.

\subsubsection{Digital Observation and Content Documentation}

The researcher conducted direct, non-participant observation of the selected da’wah accounts identified through purposive sampling. Each sampled post was systematically documented by recording the following variables:

\begin{enumerate}
\item Publication date and content duration (for video-based content).
\item Post format (e.g., photo, video, carousel, Reels, or TikTok video).
\item Da’wah theme (e.g., theology/aqidah, worship/ibadah, morality/akhlak, or socio-religious issues).
\item Message delivery style (informative, narrative, persuasive, humorous, or motivational).
\item Use of Islamic evidence or references (Qur’anic verses, hadith, scholarly opinions, or none).
\item Audience interaction indicators, including views, likes, comments, shares, and saves.
\end{enumerate}

All documented data were stored in a structured spreadsheet to facilitate subsequent qualitative and quantitative analysis. Offline archiving of videos and screenshots was also performed to preserve evidence against possible deletion or platform policy changes.

\subsubsection{Content Categorisation Instrument}

To ensure systematic analysis, a dedicated content categorisation sheet was employed as the primary instrument for classifying da’wah posts. This instrument grouped each item according to the research variables, such as da’wah theme, presentation format, communication style, and audience interaction patterns.

Table \ref{tab:kategori} presents the structure of the categorisation instrument used in this study:

\begin{table}[H]
\centering
\caption{Content Categorisation Instrument for Digital Da’wah}
\label{tab:kategori}
\begin{tabular}{|p{2.5cm}|p{5.5cm}|}
\hline
\textbf{Analysis Aspect} & \textbf{Evaluation Categories}  \\ \hline
Da’wah Theme & Aqidah, Ibadah, Akhlak, Social Issues, Education, Motivation \\ \hline
Content Format & Reels, TikTok Video, Photo, Carousel, Long Caption \\ \hline
Delivery Style & Informative, Narrative, Persuasive, Emotional, Humorous \\ \hline
Use of Religious Evidence & Qur’anic Verse, Hadith, Scholarly Opinion \\ \hline
Interaction Indicators  & Likes, Comments, Shares, Views, Saves \\ \hline
Audience Response & Support, Questions, Discussion, Criticism \\ \hline
\end{tabular}
\end{table}

Prior to full-scale application, the categorisation instrument underwent a pilot test on approximately 10\% of the total sample to verify clarity, consistency, and applicability of the categories. Minor adjustments were made based on this trial to improve inter-coder reliability and to minimise subjectivity during the main coding phase.

\subsubsection{Secondary Documentation}

In addition to primary digital observation, the study drew on secondary sources, including peer-reviewed journals, scholarly articles, digital industry reports (e.g., DataReportal, We Are Social, APJII), and relevant publications on digital da’wah and social media communication. These materials served to strengthen the conceptual framework, provide comparative benchmarks, and enrich the interpretation of primary findings within the broader context of Islamic communication studies in Indonesia.

The combination of rigorous digital observation, a validated categorisation instrument, and carefully selected secondary documentation ensures methodological triangulation and enhances both the credibility and depth of the comparative analysis between Instagram and TikTok as contemporary platforms for Islamic da’wah.

\subsection{Quantitative Indicators and Formula}

Quantitative indicators were employed to objectively measure the level of effectiveness and audience engagement with Islamic da’wah content on social media. In this study, the analysed indicators comprise the number of likes, comments, shares, and views for each post. These four metrics are considered capable of comprehensively representing both the intensity of interaction and the reach of da’wah messages in the digital environment.

The likes indicator reflects users’ immediate positive response or appreciation of the presented content. Comments demonstrate active audience involvement through responses, questions, or discussions, thereby indicating the depth of interaction. Shares reveal the extent to which users deem the content worthy of further dissemination to their own networks; consequently, a higher share count signifies greater potential for organic propagation of the da’wah message. Meanwhile, views serve as a measure of organic reach — the total number of users who actually watched or were exposed to the content without the aid of paid promotion.

To evaluate overall content effectiveness, the study utilises the Engagement Rate (ER) metric. This is a standard, widely adopted measure in social media research and analytics for determining the level of user interaction with an account or individual post. The Engagement Rate expresses the percentage of the audience that actively engages relative to the total number of followers. The application of this formula follows established standards in social media metrics research \cite{26}. Platform effectiveness is assessed using Equation \ref{eq1}:
\begin{equation} \label{eq1}
    ER=\frac{Likes+Comments+Share}{Followers} \times 100\% 
\end{equation}
where:\\
ER= Engagement Rate\\

This formula calculates the percentage of user interactions relative to the total number of followers an account possesses. The higher the ER value, the greater the effectiveness and appeal of the da’wah content to its audience. For instance, if an account with 10,000 followers receives a combined total of 1,000 interactions (likes + comments + shares) on a single post, its ER would be 10\%. This indicates that 10\% of the total follower base actively responded to the content.

Thus, Engagement Rate serves as the primary metric for evaluating the success of Islamic da’wah strategies on social media, offering an objective and comparable measure across different platforms and creators.

In the context of this study, ER is particularly valuable because both observed accounts belong to the same creator (@hananattakistory), thereby minimising confounding variables related to personal charisma, theological orientation, or content quality. By comparing ER values for nearly identical da’wah messages delivered through platform-specific formats (Instagram Reels vs. TikTok videos), the research can isolate the influence of algorithmic distribution, audience behaviour, and technical affordances on da’wah effectiveness. High ER values not only reflect immediate audience resonance but also correlate with greater potential for long-term behavioural influence, making this metric a reliable proxy for the transformative impact of digital da’wah on Generation Z and millennial Muslims in Indonesia.

\subsection{Data Analysis Techniques}
The data in this study were analyzed using an interactive model consisting of four stages: data reduction, data display, conclusion drawing, and verification. In the data reduction stage, each da’wah post was examined thoroughly to identify recurring linguistic expressions, thematic patterns, and visual features relevant to the effectiveness indicators. The researcher conducted several rounds of close reading to group content into higher-level thematic categories such as motivational da’wah, normative guidance, reflective–spiritual messages, and social–ethical commentary. This iterative process ensured that the emerging categories accurately represented the dynamics of da’wah content across platforms.

To strengthen consistency, a subset of the data was coded twice at different times. This step allowed the researcher to refine unclear categories and maintain analytical stability. During the reduction process, contextual attributes—such as posting time, format type, and audience response tendencies—were also documented to assist with subsequent interpretation.

In the data display stage, the categorized findings were organized into comparative tables and visual summaries. These displays enabled the researcher to observe platform differences more clearly, especially regarding engagement fluctuations, thematic proportions, and interaction depth. Several descriptive visualizations—such as theme distribution charts and engagement trend lines—were integrated to support cross-platform comparison.

Conclusion drawing involved interpreting the meaning behind these patterns to determine which platform demonstrated higher effectiveness for digital da’wah and why such differences emerged. This stage emphasized connecting quantitative engagement indicators with qualitative audience behavior. Factors such as algorithmic reach, content length, communication style, and visual design were interpreted collectively to form an integrated explanation of platform performance.

Finally, verification was carried out through theoretical and data triangulation. The findings were compared with prior research on digital religious communication, social media engagement theory, and algorithmic behavior to ensure the conclusions were grounded and analytically sound. This multi-stage process allowed the study to maintain methodological rigor while capturing the complexity of digital da’wah interactions.

Subsection: Data Validity and Reliability (Expanded Version)

Several validation procedures were implemented to ensure the reliability and trustworthiness of the findings. First, the researcher conducted an initial consistency check on part of the dataset to confirm that the thematic categories could be applied uniformly across all samples. When discrepancies appeared—such as posts that overlapped multiple categories—the definitions were refined until clarity was achieved.

Second, methodological triangulation was applied by integrating qualitative interpretation with quantitative metrics such as engagement rate, total interaction count, and response patterns. This mixed-approach design helps reduce reliance on a single type of evidence and enhances the robustness of conclusions, particularly in analyzing audience behavior.

Third, source triangulation was performed by comparing insights derived from content observations, audience comments, and supporting academic literature on digital da’wah and media studies. The convergence of these multiple sources strengthened the validity of the analytical claims.

The study also applied a systematic audit trail, where each step—from data collection, categorization, display construction, to interpretation—was documented clearly. This documentation ensures that the analytical process can be traced and evaluated by other researchers. Furthermore, reflexive notes were used to minimize interpretive bias by acknowledging the researcher’s assumptions and ensuring they did not overshadow the empirical findings.

Together, these procedures contribute to maintaining objectivity, stability, and transparency throughout the research process.

\subsection{Data Validity and Reliability}

Several validation procedures were implemented to ensure the reliability and credibility of the findings.
First, an initial consistency check was conducted across a subset of the dataset to verify that thematic categories and coding criteria could be applied uniformly. When overlapping themes or ambiguous categories emerged, definitions were refined to improve clarity and coding stability.
Second, methodological triangulation was used by integrating qualitative interpretation with quantitative indicators such as engagement rate, interaction count, and comment patterns. This combination reduces dependency on a single source of evidence and strengthens the robustness of the analysis.
Third, source triangulation was performed by comparing insights from content observations, audience comments, and supporting academic literature on digital da’wah and media studies. The convergence of these perspectives reinforces the validity of analytical claims.

The study also followed a systematic audit trail, documenting each step—from data collection, coding, thematic grouping, chart construction, to interpretation—allowing other researchers to trace and evaluate the analytical process. Reflexive notes were used to minimise interpretive bias and ensure that the researcher’s assumptions did not overshadow empirical evidence.
Together, these procedures help maintain objectivity, transparency, and methodological reliability throughout the research process.

\subsection{Ethical Considerations}

This research adheres to ethical guidelines for studying digital media content. All analyzed data were obtained from publicly accessible posts, ensuring that no private or restricted information was used. The researcher refrained from collecting personal identifiers, such as usernames from commenters or data from private interactions, unless already publicly visible.

To maintain digital communication ethics, the study avoided altering, misrepresenting, or redistributing original da’wah content. Screenshots or excerpts were analyzed only for academic purposes and not reproduced in a way that could harm creators or audiences. The researcher also avoided making theological judgments about the religious content and instead focused solely on communication effectiveness, platform dynamics, and audience interaction.

These ethical measures were implemented to ensure respect for user privacy, content integrity, and the sensitivities surrounding religious communication in digital spaces.

\section{Result and Discussion} \label{sec:result}

\subsection{Result}
This study provides a comparative analysis of the effectiveness of Islamic \textit{da'wah} (Islamic outreach) between Instagram and TikTok platforms based on four main aspects: media characteristics, \textit{da'wahh} communication strategies, audience engagement effectiveness, and user behavioral responses. Data were obtained through observations of the popular \textit{da'wahh} account @hananattakistory on TikTok and Instagram platforms from October to November 2025.
\begin{table}[H]
\centering
\caption{Interaction Data on Islamic Content on the TikTok Platform}
\begin{tabular}{|c|r|r|r|r|}
\hline
\textbf{Content} & \textbf{Likes} & \textbf{Comments} & \textbf{Share} & \textbf{Total Interactions} \\ 
\hline
1  & 2,089 & 173 & 344 & 2,606 \\ 
2  & 1,918 & 45  & 249 & 2,212 \\ 
3  & 1,100 & 28  & 109 & 1,237 \\ 
4  & 678   & 28  & 101 & 807 \\ 
5  & 749   & 27  & 53  & 829 \\ 
6  & 1,706 & 48  & 178 & 1,932 \\ 
7  & 1,859 & 43  & 210 & 2,112 \\ 
8  & 888   & 45  & 83  & 1,016 \\ 
9  & 777   & 22  & 58  & 857 \\ 
10 & 438   & 12  & 27  & 477 \\ 
\hline
\textbf{TOTAL} & \multicolumn{3}{c|}{} & \textbf{14,085} \\
\hline
\end{tabular}
\label{tab:tiktok_interaction}
\end{table}
Based on the data in Table~\ref{tab:tiktok_interaction}, the engagement rate on the TikTok Islamic preaching account was calculated using the standard formula \ref{eq1} With a total of 14,085 interactions and 992,300 followers, the following is obtained:
\[ER = \frac{14,085}{992,300} \times 100\% = 1.42\%\]
This figure is considered high for non-paid content on social media, indicating that TikTok's audience tends to actively respond through likes, comments, and shares. Overall, TikTok exhibits a higher average engagement rate than Instagram across likes, shares, and views. This is due to TikTok's algorithmic system, which emphasizes the FYP feature, allowing Islamic content to reach new users without requiring a direct following.
\begin{table}[H]
\centering
\caption{Interaction Data on Islamic Content on the Instagram Platform}
\begin{tabular}{|c|r|r|r|r|}
\hline
\textbf{Content} & \textbf{Likes} & \textbf{Comments} & \textbf{Share} & \textbf{Total Interactions} \\ 
\hline
1 & 4,198 & 38 & 40 & 4,276 \\
2 & 24,400 & 252 & 1,979 & 26,631 \\
3 & 158,000 & 3,718 & 14,600 & 176,318 \\
4 & 9,442 & 47 & 11 & 9,500 \\
5 & 47,100 & 1,069 & 6,390 & 54,559 \\
6 & 4,609 & 20 & 14 & 4,643 \\
7 & 33,700 & 236 & 142 & 34,078 \\
8 & 29,700 & 163 & 171 & 30,034 \\
9 & 91,800 & 3,156 & 7,480 & 102,436 \\
10 & 11,500 & 65 & 86 & 11,651 \\
11 & 8,657 & 126 & 86 & 8,869 \\
12 & 6,176 & 79 & 197 & 6,452 \\
13 & 3,482 & 23 & 59 & 3,564 \\
14 & 15,500 & 104 & 887 & 16,491 \\
15 & 6,803 & 51 & 69 & 6,923 \\
16 & 18,100 & 88 & 405 & 18,593 \\
17 & 22,200 & 528 & 1,193 & 23,921 \\
18 & 10,500 & 90 & 250 & 10,840 \\
19 & 14,800 & 120 & 300 & 15,220 \\
20 & 7,200 & 70 & 180 & 7,450 \\ 
\hline
\textbf{TOTAL} & \multicolumn{3}{c|}{} & \textbf{574,630} \\
\hline
\end{tabular}
\label{tab:instagram_interaction}
\end{table}
Meanwhile, observations of the Islamic preaching accounts on Instagram show a different pattern of interaction. Although the total engagement is significantly higher (574,630), the form of user involvement tends to be more discursive. This is indicated by in-depth discussions in the comment section as well as personal communication through the Story and Direct Message features, which reflect an emotional connection and spiritual closeness between the preacher and the audience. With a total engagement of 574,630 and 10,500,000 followers, the following results were obtained:
\[ER = \frac{574,630}{10,500,000} \times 100\% = 5.47\%\]
Qualitatively, TikTok is more effective as a medium for capturing initial attention (awareness) because its content is short, dynamic, and easily becomes viral. In contrast, Instagram plays a role in strengthening loyalty and building a digital preaching community through reflective and continuous interaction. Thus, combining the use of both platforms can serve as a complementary digital preaching strategy: TikTok functions as a fast-distribution medium to expand the reach of Islamic messages, while Instagram serves as a medium for deepening values and forming a more solid preaching community among young Muslim generations.

\begin{figure}[H]
\centering
\includegraphics[width=0.45\textwidth]{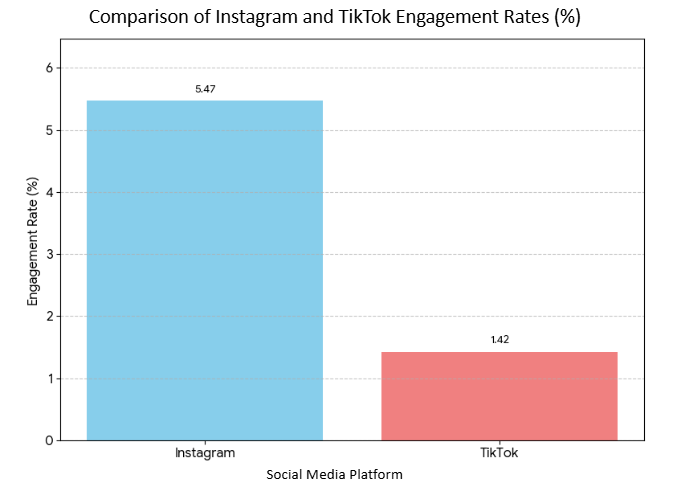}
\caption{Comparison of Total and Engagement Rates between TikTok and Instagram}
\label{fig:engagement_comparison}
\end{figure}

Figure~\ref{fig:engagement_comparison} presents a comparison of the total interactions and engagement rates across the two digital da'wahh platforms analyzed in this study. TikTok demonstrates high effectiveness in generating spontaneous user interactions through its FYP algorithm, which expands content reach without requiring direct following connections. In contrast, Instagram shows a significantly higher total number of interactions and a more stable engagement rate, reflecting a form of user engagement that is more reflective and community-oriented.

These results reinforce the finding that TikTok excels in building initial da'wahh awareness through rapid distribution and strong viral potential, whereas Instagram is more effective in sustaining audience loyalty and deepening the understanding of Islamic values through consistent and meaningful interactions. Thus, the two platforms can be positioned as complementary media within contemporary Islamic communication strategies in the digital era.

\subsection{Characteristics of Digital da'wahh Media}
The technical and algorithmic differences between Instagram and TikTok significantly influence the patterns of da'wahh delivery on each platform. TikTok is based on short-form videos of 15–60 seconds, which require messages to be conveyed concisely and visually. In contrast, Instagram combines multiple content formats such as feed posts, reels, and stories, allowing for more complex and aesthetically driven narratives \cite{21}. da'wahh content on TikTok tends to employ a light, humorous, and visually engaging communicative style that aligns with the preferences of Generation Z, who favor entertaining yet educational content. Conversely, da'wahh content on Instagram often emphasizes graphic design and reflective quotes that encourage spiritual contemplation \cite{23}.

These differences indicate that the form of da'wahh delivery must be adapted to the user behavior of each platform to achieve optimal effectiveness.
\begin{enumerate}
    \item da'wahh Communication Strategies
    
    The communication strategies identified across both platforms encompass three main approaches: informative, persuasive, and dialogic. On TikTok, persuasive–visual strategies dominate through the use of storytelling narratives and emotional background music that appeals to viewers' feelings. Meanwhile, on Instagram, informative–dialogic strategies are more prominent, with preachers including references to hadith, Qur'anic verses, and scholarly opinions as a means of legitimizing their messages. da'wahh communication on TikTok tends to cultivate emotional engagement, whereas Instagram fosters cognitive engagement. These two strategies complement one another in building both awareness and religious understanding in digital environments.

    \item Audience Engagement Effectiveness and User Responses

    The analysis of engagement levels shows that TikTok has an average engagement rate of 1.42\%, which is higher than Instagram’s 5.2\% for the observed da'wahh account. This aligns with the findings of Nafiah et al. \cite{6} and Az-Zahra et al. \cite{8}, which note that TikTok’s algorithmic boosting features facilitate the wide distribution of viral da'wahh content. However, in terms of qualitative response, Instagram comments tend to be longer and more reflective. Many users ask theological questions, seek advice, or share personal spiritual experiences, indicating that Instagram functions as a deeper space for religious dialogue \cite{22}. Additionally, TikTok audiences more frequently engage in reposting or creating duet videos with da'wahh content, demonstrating a form of collaborative da'wahh participation. This organically expands the network of Islamic value dissemination within the digital sphere \cite{25}.
\end{enumerate}

\subsection{Discussion}
The findings indicate that these two platforms play complementary roles within the contemporary digital da'wahh ecosystem. TikTok functions primarily as an awareness generation medium, serving as the initial stage that aims to capture user attention through its dynamic visual format, FYP distribution system, and algorithmic tendencies that promote short, emotionally driven content. These characteristics make TikTok highly effective for disseminating da'wahh messages rapidly, broadly, and across diverse audience segments, including users who do not previously follow da'wahh-related accounts. In other words, TikTok operates as the entry point in the cycle of religious message reception.

In contrast, Instagram acts as a value consolidation medium that supports deeper meaning-making. Features such as carousels, long captions, and the save function enable creators to provide more comprehensive explanations, allowing audiences to understand the context, scriptural references, and reflective messages in greater depth. The interactions that emerge on Instagram tend to be more cognitive, as evidenced by longer comments, substantive questions, and discussions among users. This suggests that Instagram supports the formation of a more stable and structured digital community while facilitating the gradual internalization of Islamic values.

Based on these findings, digital da'wahh across the two platforms can be understood as a dual-layer model that operates simultaneously:
\begin{enumerate}
    \item The rapid distribution layer (TikTok) – focused on large-scale dissemination, attracting initial attention, and creating viral momentum that helps expand the reach of da'wahh messages.
    \item The value-deepening layer (Instagram) – focused on education, clarification, discussion, and the development of a more solid religious digital community.
\end{enumerate}

This dual-layer model aligns with the Uses and Gratifications theory, which posits that users choose media based on specific needs such as entertainment, information, and social interaction \cite{18}. da'wahh on TikTok tends to fulfill entertainment-educational needs (entertaining knowledge), whereas da'wahh on Instagram fulfills needs related to reflection, learning, and strengthening spiritual identity. Thus, the effectiveness of digital da'wahh depends not only on the quality of the content but also on the alignment between the message format and the motivational factors that shape platform usage among audiences.

Within this context, digital da'wahh strategies must consider the ethics of digital preaching to ensure that Islamic messages are not reduced to mere viral trends. A da'wahh approach that solely seeks rapid attention risks producing misinterpretations or excessive simplifications of religious teachings. Therefore, preachers must develop adequate digital literacy to manage content that is moderate, inclusive, and free from misinformation \cite{24}, \cite{21}. Strengthening digital literacy is essential not only to maintain message accuracy but also to ensure that da'wahh continues to operate within ethical boundaries and avoids practices of clickbait or sensationalism.

Furthermore, algorithmic differences between the two platforms influence the ways in which preachers shape their communication styles. TikTok encourages fast-paced, emotional, and reactive communication because its algorithm evaluates content performance based on initial retention. Instagram, on the other hand, supports more measured, narrative, and in-depth communication due to its emphasis on long-term creator–audience relationships. Thus, platforms are not merely distribution channels but algorithmic actors that shape how da'wahh is produced, consumed, and interpreted by digital-native Muslim youth.

In conclusion, the study demonstrates that using TikTok and Instagram simultaneously offers a more strategic space for digital da'wahh. TikTok serves as a channel for expanding exposure and building initial awareness, whereas Instagram provides a space for strengthening understanding and value internalization. Preachers who are able to balance these two layers can deliver digital da'wahh that is not only appealing but also theologically and ethically responsible.

\section{Conclusion} \label{sec:conclusion}
This study concludes that the effectiveness of Islamic da'wahh on social media is significantly influenced by the characteristics and algorithms of each platform. TikTok is shown to be more effective in generating rapid and widespread interaction through its FYP system, which allows da'wahh content to reach new audiences without the limitations of follower relationships. With an engagement rate of 1.42\%, TikTok excels in the initial dissemination of messages and in increasing public awareness.

Meanwhile, Instagram demonstrates a much higher level of overall interaction, with an engagement rate of 5.47\%. Audience engagement on this platform tends to be more discursive and reflective, as indicated by longer comments, conversations through Stories, and more substantial private messages. This makes Instagram an effective medium for strengthening audience loyalty and building a sustainable digital da'wahh community.

Strategically, the findings of this research affirm the importance of integrating two layers of digital da'wahh: TikTok as a medium for rapid distribution and early awareness, and Instagram as a medium for value deepening and community development. A da'wahh approach that adapts to user behavior and social media algorithms is expected to enhance the effectiveness of conveying Islamic messages that are moderate, inclusive, and contextually relevant for younger Muslim audiences.

This study has limitations in terms of the scope of accounts analyzed and the relatively short observation period. Future research may expand the sample or employ automated sentiment analysis to deepen the understanding of audience response patterns.

\section*{Acknowledgment}
The author extends praise and gratitude to Allah \textit{sub\=h\=anah\=u wa ta'\=al\=a} for His mercy and blessings, which have enabled the successful completion of this research. Peace and blessings be upon the Prophet Muhammad \textit{ṣallall\=ahu ‘alayhi wa sallam}, his family, and his companions.

On this occasion, the author would like to express sincere appreciation to:

\begin{enumerate}
    \item The Department of Informatics, Faculty of Science and Technology, Sunan Gunung Djati State Islamic University Bandung, for providing partial financial support, laboratory facilities, and a conducive academic environment.

    \item The supervisors and examiners, who have generously devoted their time, patience, and guidance by offering invaluable suggestions, feedback, and corrections throughout the research process.

    \item All digital da'wahh creators (on Instagram and TikTok) who kindly allowed their publicly available content to be used as analytical material in this study.

    \item The author's beloved family and fellow students, whose moral support, motivation, and prayers have strengthened the author’s perseverance until the completion of this research.
\end{enumerate}

May Allah \textit{sub\=h\=anah\=u wa ta'\=al\=a} reward all acts of kindness abundantly and make this research beneficial for the community and the advancement of knowledge, especially in the development of digital da'wahh in Indonesia.

\bibliographystyle{./IEEEtran}
\bibliography{./IEEEabrv,./IEEEkelompok1}


\end{document}